\documentstyle[12pt,titlepage]{article} 
\title{The Decay $\eta_c \rightarrow \gamma \gamma$ : A Test for
  Potential Models}
\author{Mohammad R. Ahmady and Roberto R. Mendel}          
\date{January, 1994}   
\def\_{\rule{.3em}{.15ex}}  

\begin{document}           
\begin{titlepage}
\begin{flushright}
 HEP-PH 9401315
\end{flushright}
 \begin{center}
  \vspace{0.75in}
  {\bf {\LARGE The Decay $\eta_c \rightarrow \gamma \gamma$ : A Test
for
  Potential Models} \\
  \vspace{0.75in}
  Mohammad R. Ahmady and Roberto R. Mendel} \\
  Department of Applied Mathematics\\
  The University of Western Ontario \\
  London, Ontario, Canada\\
  \vspace{0.75in}
  ABSTRACT \\
  \vspace{0.35in}
  \end{center}
  \begin{quotation}
\begin{sloppypar}
\noindent We use a simple perturbation theory argument and
measurements of
  charmonium leptonic widths $\Gamma (\psi_{NS} \rightarrow e^+e^-)$
to
 estimate the ratio \mbox{$R_\circ \equiv  {\vert \Psi
_{\eta_{c1S}}(0) \vert}^2
/{\vert
 \Psi_{\psi_{1 S}}(0)\vert}^2$} in the general context of non-
relativistic potential
 models.  We obtain $R_\circ = 1.4 \pm 0.1$.  We then apply well
known
 potential model formulas, which include lowest order QCD
corrections, to find
 $\Gamma (\eta_c \rightarrow \gamma \gamma )/\Gamma (\psi_{1S}
\rightarrow e^+
 e^-) \approx 2.2\pm 0.2$.  The central value for $\Gamma (\psi_{1S}
 \rightarrow e^+ e^-)$in the 1992 Particle Data Tables then leads to
a
 (non relativistic) prediction $\Gamma (\eta_c \rightarrow \gamma
\gamma )\approx 11.8\pm
0.8 $
 keV.  This prediction is in good agreement with a recent measurement
by
 the ARGUS collaboration, is consistent with a recent measurement by
the
 L3 collaboration but is significantly higher than several earlier
measurements
 and than previous theoretical estimates, which usually assume
$R_\circ =1$.
The correction to $R_\circ =1$ is estimated to be smaller but
nonnegligible for the $b\bar b$ system.  Using the current central
measurement for $\Gamma (\Upsilon_{1S}\rightarrow e^+e^-)$ we find
$\Gamma (\eta_b\rightarrow \gamma \gamma )\approx 0.58\pm 0.03$ keV.
A rough estimate of relativistic corrections reduces the expected two
photon rates to about 8.8 keV and 0.52 keV for the $\eta_c$ and
$\eta_b$ mesons respectively.  Such corrections however, are not
expected to significantly affect our estimates of $R_\circ$.

\end{sloppypar}
\end{quotation}
\end{titlepage}


\newcommand{\da}{\mbox{$\scriptscriptstyle \dag$}}
\newcommand{\lag}{\mbox{$\cal L$}}
\newcommand{\tr}{\mbox{\rm Tr\space}}
\newcommand{\fc}{\mbox{${\widetilde F}_\pi ^2$}}
\newcommand{\ns}{\textstyle}
\newcommand{\si}{\scriptstyle}

Potential models (non-relativistic [NR] as well as `` relativized''
versions) have been successfully used to describe many properties of
quarkonium
($c\bar c$ and $b\bar b$) states \cite {PM}.  Relativistic effects
and
beyond lowest order QCD corrections are expected to be more important
in
$c\bar c$ meson than in $b\bar b$ meson but it is hard to devise
model
independent tests (i.e. that do not depend strongly on the particular
form
of the potential being used) that may pinpoint some properties of
charmonium
spectroscopy or decays where a N-R description clearly fails.

In the present paper we make a prediction for $\Gamma (\eta_c
\rightarrow \gamma \gamma )$ $(\Gamma (\eta_b \rightarrow \gamma
\gamma ))$
which relies solely on i) a N-R description of the $c \bar c$ $(b\bar
b)$ system,
 ii) experimental data of leptonic charmonium (bottomium) decays,
iii)
approximate validity of lowest order perturbation theory for the
color-hyperfine splitting interaction and iv) approximate validity of
lowest
order QCD radiative corrections.  At the end of the paper we make an
estimate of relativistic corrections to our results.

As a starting point we make use of a previously derived result \cite
{KMR}
relying on a N-R description of quarkonium systems which
includes lowest order QCD corrections:
\begin{equation}
\frac {\Gamma (\eta_c \rightarrow \gamma \gamma )}{\Gamma (\psi_{1S}
\rightarrow e^+e^-)}=\frac {4}{3} (1+1.96 \frac {\alpha_s}{\pi})\frac
{
M_\psi ^2}{{(2m_c)}^2}\frac {{\vert \Psi _{\eta_c}(0) \vert}^2}{{\vert
 \Psi_{\psi_{1 S}}(0)\vert}^2}\; \; ,
 \end{equation}
where $\alpha_s$ should be evaluated at the charm scale.  We will use
$
\alpha_s(m_c)\approx 0.28\pm 0.02$ \cite {KMR}.  For consistency of
the
N-R description of the system and since the hyperfine splitting will
be included only perturbatively, one should set $2m_c\approx M_\psi$
\cite {BU}.

 Thus, one obtains
\begin{equation}
\frac {\Gamma (\eta_c \rightarrow \gamma \gamma )}{\Gamma (\psi_{1S}
\rightarrow e^+e^-)}\approx 1.57 \frac {{\vert \Psi _{\eta_c}(0)
\vert}^2}{{\vert
 \Psi_{\psi_{1 S}}(0)\vert}^2}\;\; ,
 \end{equation}
and analogously, using $\alpha_s (m_b)\approx 0.18 \pm 0.01$ \cite
{KMR}
\begin{equation}
\frac {\Gamma (\eta_b \rightarrow \gamma \gamma )}{\Gamma
(\Upsilon_{1 S}
\rightarrow e^+e^-)} \approx 0.37 \frac {{\vert \Psi _{\eta_b}(0)
\vert}^2}{{\vert
 \Psi_{\Upsilon_{1 S}}(0)\vert}^2}\;\; .
 \end{equation}

It is important to note that even if we were far more conservative
with
the uncertainty in the value of $\alpha_s(m_c)$ and $\alpha_s(m_b)$,
the
resulting uncertainty in the numerical coefficients of eqns. (2) and
(3) would
only be of order of a few percent.

A widely used approximation at this point is to set ${\vert
\Psi _{\eta_{c}}(0) \vert}^2/{\vert \Psi_{\psi_{1 S}}(0)\vert}^2
\approx
1$\\ $
 ({\vert \Psi _{\eta_{b}}(0) \vert}^2/{\vert
 \Psi_{\Upsilon_{1 S}}(0)\vert}^2 \approx 1)$ leading to $ \Gamma
 (\eta_c \rightarrow \gamma \gamma )/\Gamma (\psi_{1S} \rightarrow
e^+e^-)
 \approx 1.6$\\ $ (\Gamma (\eta_b \rightarrow \gamma \gamma )/\Gamma
 (\Upsilon_{1S} \rightarrow e^+e^-)\approx 0.37)$.  We will
show below that for the $c\bar c$ case this commonly used assumption
is off
by more than $30\%$ (and could be off by as much as $50\%$).  For the
$b\bar b$ case the correction to this approximation is estimated to be
smaller but still significant.

In the context of N-R potential models, the interaction Hamiltonian
responsible
 for the splitting between the $1^3S_1$ states and the $1^1S_0$ state
of a $Q\bar Q$
 meson is given by $\hat H=\hat H_{S_{12}}+\hat H_{SO}$ where
 \begin{equation}
 \hat H_{S_{12}} = \frac {3\vec{\mu_1}.\hat r\vec{\mu_2}.\hat
r-\vec{\mu_1}.\vec{\mu_2}}{r^3}+
 \frac {8\pi}{3}\vec{\mu_1}.\vec{\mu_2}\delta^3(\vec{r})=\hat
H^r_{S_{12}}+\hat H^\delta _{
 S_{12}}\;\; .
 \end{equation}
$\vec{\mu_1}$ and $\vec{\mu_2}$ are respectively the static color-
magnetic
moment operator for the quark and the antiquark and $\vec{r}$ is the
relative
position vector.  Color indices have been omitted and electromagnetic
interactions have been ignored. The Spin-Orbit part of the
Hamiltonian, $\hat H_{SO}$, is included for completeness.  However,
it does not contribute to lowest order to either the energy splitting
(eqn. (5) below) or the shifts of the wavefunctions (eqns. (6-9)
below) for S states.

We now study the effects of the Hamiltonian $\hat H$ to {\it lowest}
order in
perturbation theory.
The energy splitting between the $1^3S_1$ and $1^1S_0$ states is
given, to lowest
order by
\begin{eqnarray}
\Delta E^{(1)} & = & E^{(1)}_{1^3S_1}-E^{(1)}_{1^1S_0}=
<\Psi_{1^3S_1}^{(0)}\vert \hat H_{S_{12}}\vert \Psi_{1^3S_1}^{(0)}>
-<\Psi_{1^1S_0}^{(0)}\vert \hat H_{S_{12}}\vert \Psi_{1^1S_0}^{(0)}>
\nonumber \\
& = & <\Psi_{1^3S_1}^{(0)}\vert \hat H^\delta_{S_{12}}\vert
\Psi_{1^3S_1}^{(0)}>
-<\Psi_{1^1S_0}^{(0)}\vert \hat H^\delta_{S_{12}}\vert
\Psi_{1^1S_0}^{(0)}>\;\; ,
\end{eqnarray}
where the last step follows after angular integration from the fact
that the
relevant states have $\ell =0$.  Experimentally, $\Delta E_{c\bar c}
=M_{\psi_{1S}}-M_{\eta_{c1S}}=118\pm 2$  MeV \cite {PDG}.  We estimate
$\Delta E_{b\bar b}=M_{\Upsilon_{1S}} -M_{\eta_{b1S}}=45\pm 15$  MeV,
using
the measured value of $\Delta E_{c\bar c}$ and N-R potential model
formulas which include lowest order QCD corrections.

The lowest order correction to the radial and orbital ground state
wavefunctions at a point $\vec{r}$ due to the Hamiltonian $\hat H$ is
given (in common bra-ket notation) by
\begin{equation}
\Delta^{(1)}\Psi_{1 ^3S_1}(\vec{r})=\Delta^{(1)} < \vec{r} \vert
\Psi_{1 ^3S_1}>
=\sum _{\stackrel {n}{E^{(0)}_n\neq E^{(0)}_{1S}}}\frac {<\vec{r}
\vert \Psi^{(0)}_n>
<\Psi^{(0)}_n\vert \hat H_{S_{12}}\vert
\Psi^{(0)}_{1 ^3S_1}>}{
E^{(0)}_{1S}-E_n^{(0)}}\;\; ,
\end{equation}
\begin{equation}
\Delta^{(1)}\Psi_{1 ^1S_0}(\vec{r})=\Delta^{(1)} < \vec{r} \vert
\Psi_{1 ^1S_0}>
=\sum _{\stackrel {n}{E^{(0)}_n\neq E^{(0)}_{1S}}}\frac {<\vec{r}
\vert
\Psi^{(0)}_n><\Psi^{(0)}_n\vert \hat H_{S_{12}}\vert
\Psi^{(0)}_{1 ^1S_0}>}{
E^{(0)}_{1S}-E_n^{(0)}}\;\; .
\end{equation}

But the wavefunction at the origin is zero for all non-S states and
therefore
only S-states contribute to the sum for $\vec{r} =0$.  Thus, the {\it
shifts
in the wavefunctions at the origin} are given as follows:
\begin{eqnarray}
\Delta^{(1)}\Psi_{1 ^3S_1}(\vec{0})
& = & \sum _{N>1}\frac {< \vec{0} \vert
\Psi^{(0)}_{NS}><\Psi^{(0)}_{NS}\vert \hat H_{S_{12}}\vert
\Psi^{(0)}_{1 ^3S_1}>}{
E^{(0)}_{1S}-E_{NS}^{(0)}} \nonumber \\
& = & \sum _{N>1}\frac {< \vec{0} \vert
\Psi^{(0)}_{NS}><\Psi^{(0)}_{NS}\vert \hat H^\delta
_{S_{12}}\vert \Psi^{(0)}_{1 ^3S_1}>}{
E^{(0)}_{1S}-E_{NS}^{(0)}}  \nonumber \\
& = & \sum _{N>1}\frac {{\vert
\Psi^{(0)}_{NS}(\vec{0})\vert}^2<\Psi^{(0)}_{1
^3S_1}\vert
\hat H^\delta_{S_{12}}\vert \Psi^{(0)}_{1 ^3S_1}>}{(E^{(0)}_{1S}-
E_{NS}^{(0)})
{\Psi^{(0)}_{1 ^3S_1}(\vec{0})}^\ast}\;\; .
\end{eqnarray}
The last step is possible because the matrix element has support only
at the
origin.  The previous step follows from the fact that only S-states
are
involved in the sum.  Note that only one state per radial excitation
contributes to the sum.  The wavefunction in the denominator is a
short hand
notation to mean that multiplication on the left by $<\Psi^{(0)}_{1
^3S_1}
\vert \vec{0}>$ will cancel it.

In the same way,
\begin{equation}
\Delta^{(1)}\Psi_{1 ^1S_0}(\vec{0})
=\sum _{N>1}\frac {{\vert
\Psi^{(0)}_{NS}(\vec{0})\vert}^2<\Psi^{(0)}_{1
^1S_0}\vert
\hat H^\delta_{S_{12}}\vert \Psi^{(0)}_{1 ^1S_0}>}{(E^{(0)}_{1S}-
E_{NS}^{(0)})
{\Psi^{(0)}_{1 ^1S_0}(\vec{0})}^\ast}\;\; .
\end{equation}

Therefore, to {\it lowest} order in perturbation theory,
\begin{eqnarray}
{\vert \Psi_{1 ^3S_1}(\vec{0})\vert}^2-
{\vert \Psi_{1 ^1S_0}(\vec{0})\vert}^2
& = & \;  \left [{\Psi^{(0)}_{1 ^3S_1}(\vec{0})}^\ast \left
(\Delta^{(1)}\Psi_{1
^3S_1}(\vec{0})\right )+
\left ({\Delta^{(1)}\Psi_{1 ^3S_1}(\vec{0})}^\ast \right
)\Psi^{(0)}_{1
^3S_1}(\vec{0})\right ] \nonumber \\ \nonumber \\
&   &\!\! \! -  \left [{\Psi^{(0)}_{1 ^1S_0}(\vec{0})}^\ast \left
(\Delta^{(1)}\Psi_{1
^1S_0}(\vec{0})\right )+
\left ({\Delta^{(1)}\Psi_{1 ^1S_0}(\vec{0})}^\ast \right
)\Psi^{(0)}_{1
^1S_0}(\vec{0})\right ] \nonumber \\
& = & 2  \Delta E^{(1)}\sum _{N>1} \frac {{\vert
\Psi^{(0)}_{NS}(\vec{0})\vert}^2}{E^{(0)}_{1S}-
E_{NS}^{(0)}}\;\; ,
\end{eqnarray}
where $\Delta E^{(1)}$ is given by eqn. (5).

Thus,
\begin{eqnarray}
\frac { {\vert \Psi _{1 ^1S_0}(\vec{0}) \vert}^2}
{{\vert\Psi_{1 ^3S_1}(\vec{0})\vert}^2}
& = & 1+
2\Delta E^{(1)}\sum _{N>1} \frac {{\vert
\Psi^{(0)}_{NS}(\vec{0})\vert}^2}
{{\vert\Psi_{1 ^3S_1}(\vec{0})\vert}^2}
\frac {1}{E_{NS}^{(0)}- E^{(0)}_{1S}} \nonumber \\
& \approx & 1+
2\left ( M_{1 ^3S_1}-M_{1 ^1S_0} \right )\sum _{N>1} \frac {{\vert
\Psi_{N
^3S_1}(\vec{0})\vert}^2}
{{\vert\Psi_{1 ^3S_1}(\vec{0})\vert}^2}
\frac {1}{M_{N ^3S_1}- M_{1 ^3S_1}}\;\; .
\end{eqnarray}
where the last step (which is consistent to lowest order) was taken
to express
the result in terms of experimentally measurable quantities.

Notice that since we are working in the context of N-R potential
models we
can express the ratio of the wavefunctions at the origin for S vector
states
in terms of ratios of leptonic widths and masses \cite {KMR}:
\begin{equation}
 \frac {{\vert \Psi_{N ^3S_1}(0)\vert}^2}
{{\vert\Psi_{1 ^3S_1}(0)\vert}^2}=
\frac {\Gamma (N ^3S_1 \rightarrow e^+e^-)M^2_{N ^3S_1}}
{\Gamma (1 ^3S_1 \rightarrow e^+e^-)M^2_{1 ^3S_1}}\;\; .
\end{equation}

Our result is then
\begin{equation}
\frac { {\vert \Psi _{1 ^1S_0}(0) \vert}^2}
{{\vert\Psi_{1 ^3S_1}(0)\vert}^2}=1+
2\left ( M_{1 ^3S_1}-M_{1 ^1S_0} \right )\sum _{N>1}
\frac {\Gamma (N ^3S_1 \rightarrow e^+e^-)M^2_{N ^3S_1}}
{\Gamma (1 ^3S_1 \rightarrow e^+e^-)M^2_{1 ^3S_1}}
\frac {1}{M_{N ^3S_1}- M_{1 ^3S_1}}\;\; .
\end{equation}

We calculate this ratio using $M_{\psi_{1S}}-M_{\eta_{c1S}}=118\pm2$
MeV as well as the information on masses and leptonic widths
 for the $^3S_1$ states
 $\psi (3097), \psi (3685), \psi (3770),  \psi (4040), \psi (4160) \\
 {\rm
 and} \; \psi (4415)$ given in the 1992 Particle Data Tables \cite
{PDG}.  We obtain
\begin{equation}
\frac { {\vert \Psi _{\eta_c 1 S}(0) \vert}^2}
{{\vert\Psi_{\psi 1 S}(0)\vert}^2}=1.4 \pm 0.1 \;\; .
\end{equation}

The error was estimated from the given experimental uncertainties
\cite {PDG}
.  We have ignored possible contributions from "continuum" states
above $D
\bar D$ threshold because these are outside the potential model
Hilbert
Space.  We are assuming implicitly that the discrete spectrum
constitutes a complete set of states of the relevant N-R
Schr\"{o}dinger equation and thus including (physical) continuum
state contributions would amount to double-counting.

We note that over half of the contribution of the sum over states in
eqn. (13) to our result in eqn. (14) is due to the lowest excitation
$\psi (3685)$ ($\approx 0.23$) while the contribution of the highest
observed radial excitation $\psi (4415)$ is quite small ($\approx
0.03$).  Therefore, we expect that the five radial excitations of
$\psi_{1S}$ observed so far saturate the sum to a good approximation.

Although we have no reliable estimate of the corrections to the
result of eqn. (14) due to effects that are higher order in $\hat
H_{S_{12}}$, the magnitude of the lowest order correction ($\approx
40\%$) can be used as an indication that such corrections are not
likely to change our result significantly (i.e. by more than
O(10\%)).  Such corrections should be viewed as a systematic
uncertainty of our approximations and will be ignored hereafter.

Combining our results in eqns. (2) and (14), we obtain the prediction
\begin{equation}
\frac {\Gamma (\eta_c \rightarrow \gamma \gamma )}{\Gamma (\psi_{1S}
\rightarrow e^+e^-)}\approx 2.2 \pm 0.2\;\; .
\end{equation}
Using the value for the $\psi_{1 S}$ leptonic width quoted in
ref.\cite {PDG},
\begin{equation}
\Gamma (\psi \rightarrow e^+e^-)=5.36 \pm ^{0.29} _{0.28} \;\;{\rm
keV},
\end{equation}
we obtain an absolute estimate
\begin{equation}
\Gamma (\eta_c \rightarrow \gamma \gamma )=11.8 \pm 0.8 \pm 0.6 \;\;
{\rm
keV},
\end{equation}
where the first uncertainty stems from our main result eqn. (15),
while the
second reflects the experimental uncertainty in eqn. (16).

We point out that if we had used the common approximation
$
{\vert \Psi _{\eta_c}(0) \vert}^2/{\vert\Psi_\psi (0)\vert}^2=1$
instead
of our estimate (eqn. (14)), we would have obtained a rate $\Gamma
(\eta_c
\rightarrow \gamma \gamma ) \approx 8.4 \pm 0.5$ keV instead of our
result in
eqn. (17).  This is in line with most previous theoretical
predictions
(around 7 keV)\cite {IS} and also agrees within errors with the
experimental
average
$(\Gamma (\eta_c \rightarrow \gamma \gamma )
=6.6 \pm ^{2.4}_{ 2.1} \; {\rm keV})$
quoted in ref. \cite {PDG} (See also refs. \cite {PLUTO}, \cite {TPC}
and \cite {CLEO}).  A more recent measurement
by the ARGUS Collaboration \cite {ARG}, on the other hand, is
\mbox{$\Gamma (\eta_c \rightarrow \gamma \gamma )
=12.6 \pm 4.0 \; {\rm keV}$}, which is centered closer to our
prediction (eqn.
(17)).
  Finally, a recently published result by L3 \cite {L3},
\mbox{$\Gamma (\eta_c \rightarrow \gamma \gamma )
=8.0 \pm 2.3 \pm 2.4 \; {\rm keV}$}, is centered closer to the
estimate based on
 the assumption $R_\circ =1$ but is still consistent with our
prediction
 (eqn. (17)).

 Although the effect is smaller for the $b\bar b$ mesons, we repeat
the
 same procedure for completeness.  Using our above mentioned estimate
 for the energy splitting $\Delta E_{b\bar b} = M_{\Upsilon_{1 S}}-
M_{\eta_{b 1S}}
 = 45\pm 15 $ MeV and the experimental information about $\Upsilon$
states
 (masses and leptonic widths) \cite {PDG}, we obtain an estimate,
  analogous to the result in eqn. (14), of
\begin{equation}
\frac {{\vert \Psi _{\eta_{b 1 S}}(0) \vert}^2}{{\vert
 \Psi_{\Upsilon_{1 S}}(0)\vert}^2}\approx 1.16 \pm 0.06\;\; .
 \end{equation}

Using our results,
 eqn. (3) and eqn. (18), we obtain
\begin{equation}
\frac {\Gamma (\eta_{b} \rightarrow \gamma \gamma )}{\Gamma
(\Upsilon_{1S}
\rightarrow e^+e^-)}\approx 0.43 \pm 0.02\;\; .
 \end{equation}

The current experimental value for $\Gamma (\Upsilon_{1S} \rightarrow
e^+e^-)$
\cite {PDG} then leads to the prediction
\begin{equation}
\Gamma (\eta_{b} \rightarrow \gamma \gamma )=0.58\pm 0.03 \pm
0.02\;\; {\rm keV}\;\; ,
\end{equation}
where the first error stems from eqn. (19) and the second one from
the
present experimental uncertainty in $\Gamma (\Upsilon_{1 S}
\rightarrow
e^+e^-)$\cite {PDG}.

To summarize, the main point of the present paper is to show that
within the
general context of N-R potential models one can (using available
experimental information) make a reliable estimate for the ratio
$R_\circ = {\vert \Psi _{1 ^1S_0}(0) \vert}^2/{\vert
 \Psi_{1 ^3S_1}(0)\vert}^2$ for $c\bar c$ and $b\bar b$ systems.  A
commonly
 used approximation is $R_\circ =1$.  Our estimates (eqns. (14) and
(18))
 imply that $R_\circ =1$ is a bad approximation for the $c\bar c$
system and
requires moderate corrections for
the $b\bar b$ system.

 The main measurable consequence of this study is that $\Gamma
(\eta_c
 \rightarrow \gamma \gamma )$ is expected to be significantly larger
 (by between $30\% {\rm and} 50\%$) than predictions based on the
$R_\circ
  =1$ assumption.  In fact, the N-R ``potential model'' predictions
for the
  rates of all important decay modes of the $\eta_c$ (relative to
$\Gamma (\psi
  \rightarrow e^+e^-)$) are enhanced by the same amount with respect
to the
  predictions based on the $R_\circ =1$ assumption.  Thus, if
$R_\circ =1$
is used for such predictions the error made would have to be
compensated
by using unphysically large values for the strong coupling $\alpha_s$
at
tree level.

{}From the theoretical viewpoint, precise measurement of
 $\Gamma (\eta_c \rightarrow \gamma \gamma )$ favouring lower values
 (eg. around 7 keV, which
 agrees well with the central value in ref. \cite {PDG} and also the
recent
 measurement ref. \cite {L3}) would put into question the validity of
N-R
 potential model assumptions for the description of $c\bar c$
systems.  The
 only way that we see to escape this conclusion would be to argue
that the
 ratio $\left ( \frac {M_\psi}{2m_c} \right )$ in eqn. (1) is
significantly
 smaller than one.  However, this seems to go against the weak-binding
 assumption that is needed for self-consistency of N-R potential
models.

 On the other hand, if $\Gamma (\eta_c \rightarrow \gamma \gamma )$
turns
 out to be close to our prediction (eqn. (17)) [See also the measured
value
 in ref. \cite {ARG}] we could state that N-R potential model
descriptions of
 charmonium have passed yet another test successfully.

All the above results and discussions are based on a strictly N-R
description of the $c\bar c$ and $b\bar b$ system.  As stated at the
beginning of the paper, relativistic corrections may be significant,
especially for the $c\bar c$ system.  We conclude our paper with a
brief discussion of the likely effects of such corrections on our
predictions.  We think that our results for the ratios $R_\circ$
(eqns. (14) and (18)) should be {\it essentially unaffected} by
relativistic corrections, because the main input for those estimates
consists of actual experimental data (leptonic widths and mass
splittings) which of course include full relativistic corrections.
The argument used for these estimates (eqns. (4) to (13)) relies more
on lowest order perturbation theory than on strictly N-R dynamics.  On
the other hand, eqn. (1) and its analogue for $b\bar b$ systems does
rely more directly on the N-R description.  The main relativistic
correction to that ratio comes from the fact that the contribution of
the quark propagator between the two photons in the $\eta_c$
($\eta_b$) decays is sensitive to the momentum distribution of the
quarks in the decaying meson [see, for example, ref. \cite {HI} and
\cite {LCB} ].  We roughly estimate these effects by using commonly
quoted values for $\frac {v^2}{c^2}|_{ave}\equiv <\beta^2>$ for the
orbital and radial ground state $c\bar c$ and $b\bar b$ quarkonium:
$<\beta^2>_{\psi} \approx 0.25$ and $<\beta^2>_\Upsilon \approx 0.1$.
The $\beta$-dependence in the integrands of eqns. (13) and (15) in
ref. \cite {HI} is `` pulled out '' of the integrals by replacing
$\beta^2$ by $<\beta^2>$.  This procedure results in a relativistic
correction factor of
$$ \frac{\frac {1-<\beta^2>}{\sqrt {<\beta^2>}}ln \left [ \frac
{1+\sqrt
{<\beta^2>}}{\sqrt {1-<\beta^2>}} \right ]}{\frac {1}{3} \left [ 2
+\sqrt
{1-<\beta^2>} \right ]}
$$ to the RHS of eqn. (1) and its $b\bar b$ analogue.  Using the
above values for $<\beta^2>$ leads to a numerical factor of about
$1\over {1.34}$ for the $c\bar c$ system and $1\over {1.11}$ for the
$b\bar b$ system.  Using our N-R central estimates given in equations
(17) and (20) above, we obtain `` relativistically corrected ''
central estimates of $\Gamma_{REL} (\eta_c \rightarrow \gamma\gamma
)\approx 8.8$ keV and $\Gamma_{REL} (\eta_b \rightarrow \gamma\gamma
)\approx 0.52$ keV.

\newpage
{\Large \bf Acknowledgments} \\ \\
We would like to thank Popat Patel for useful discussions on
experimental aspects of the
process $\eta_c\rightarrow \gamma \gamma$ and Ted Barnes for
discussions on relativistic corrections.  This work was supported
in part by the Natural Sciences and Engineering Research Council of
Canada.

\end{document}